\documentstyle[twoside,fleqn,npb,epsfig]{article}
\newcommand{\msbar}{\mbox{$\overline{\rm{MS}}$}\ }

\newcommand{\asmz}{\alpha_s(M_Z^2)}

\newcommand{\si}{\sigma}

\newcommand{\Gl}[1]{Eq.~(\ref{#1})}
\newcommand{\Ab}[1]{Fig.~\ref{#1}}

\newcommand{\gev}{GeV$^2$}

\newcommand{\beq}[1]{\begin{equation}\label{#1}}
\newcommand{\eeq}{\end{equation}}
 
\newcommand{\beqar}[1]{\begin{eqnarray}\label{#1}}
\newcommand{\eeqar}{\end{eqnarray}}

\newcommand{\AmS}{{\protect\the\textfont2
  A\kern-.1667em\lower.5ex\hbox{M}\kern-.125emS}}

\title{A QCD analysis of HERA and fixed target structure function data}

\author{M. Botje\address{NIKHEF,
        P.O. Box 41882, 1009 DB Amsterdam, The Netherlands}}

\begin{document}

\begin{abstract}
The parton momentum densities in the proton are determined from a NLO
QCD analysis of structure functions measured by HERA and fixed target
experiments. Also included are data on the difference of the up and down
anti-quark densities. The uncertainties in the parton densities,
structure functions and related cross sections are estimated from the
experimental errors, taking into account all correlations. Standard
Model predictions of the charged current Born cross sections at large
$x$ and $Q^2$ are calculated and compared with recent data from ZEUS. 
\end{abstract}

\maketitle

\section{INTRODUCTION}

With the integrated luminosity of about 50~pb$^{-1}$ collected at HERA
during the years 1994--1997 a new kinematic domain of large $x$ and
$Q^2$ becomes accessible for the study of deep inelastic scattering in
$ep$ collisions. Measurements by ZEUS of the $e^+p$ single differential
neutral current (NC) and charged current (CC) Born cross sections for
$Q^2 > 200$~\gev\ have recently become
available~\cite{ref:ncxsec,ref:ccxsec}. 

Standard Model (SM) predictions calculated with, for instance, the
parton distribution set CTEQ4~\cite{ref:cteq4} are in good agreement
with the NC cross sections but fall below the CC measurements at large
$x$ and $Q^2$. This could be an indication of new physics beyond the SM
but might also be due to an imperfect knowledge of the parton densities
in this kinematic region. For instance in~\cite{ref:bodek} it is shown
that a modification of the CTEQ4 down quark density yields SM
predictions in agreement with the CC $e^+p$ data. 

To investigate these issues we have performed a global NLO QCD
analysis~\cite{ref:michiel} of structure function data to obtain the
parton densities in the proton.  A full error analysis provides
estimates of the uncertainties in these parton densities, the structure
functions and the SM predictions of the NC and CC cross sections. 

\section{QCD ANALYSIS}

The data used in the fit are $F_2^p$ from ZEUS~\cite{ref:zeusnv} and
H1~\cite{ref:h1nv} together with $F_2^p$,
$F_2^d$~\cite{ref:e665,ref:nmc,ref:bcdms,ref:slac} and
$F_2^d/F_2^p$~\cite{ref:f2dop} from fixed target experiments. Also
included were neutrino data on $xF_3^{\nu Fe}$~\cite{ref:ccfr} for
$x > 0.1$ and data on the difference $x(\bar{d}-\bar{u})$~\cite{ref:e866}.

After cuts the data cover a kinematic range of $10^{-3} < x < 0.75$, $3
< Q^2 < 5000$~\gev\ and $W^2 > 7$~\gev. Both $F_2^d$ and $F_2^d/F_2^p$
as well as $xF_3^{\nu Fe}$ were corrected for nuclear effects using the
parameterization of~\cite{ref:gomez} for deuterium (with an assumed
uncertainty of 100\%) and that of~\cite{ref:eskola} for iron (assumed
uncertainty 50\%). 

The QCD predictions for the structure functions were obtained by solving
the QCD evolution equations in NLO in the \msbar scheme~\cite{ref:furm}.
At the scale $Q^2_0 = 4$~\gev\ the gluon density ($xg$), the sea and
valence quark densities ($xS,xu_v,xd_v$) and the difference
$x(\bar{d}-\bar{u})$ were parameterized in the standard
way~\cite{ref:mrsa} (16 free parameters).  The strange quark density was
taken to be a fraction $K_s = 0.20 (\pm 0.03)$ of the
sea~\cite{ref:ccfrs}.  The normalizations of the parton densities were
constrained such that the momentum sum rule and the valence quark
counting rules are satisfied.  The charm and bottom quarks were assumed
to be massless and were generated dynamically above the thresholds
$Q^2_c = 4 (\pm 1)$ and $Q^2_b = 30$~\gev\ respectively. The value of
the strong coupling constant was set to $\asmz = 0.118 (\pm 0.005)$. 

Higher twist contributions to $F_2^p$ and $F_2^d$ were taken into
account phenomenologically by describing these structure functions as
\begin{equation} \label{eq:htwist}
F_2^{HT} = F_2^{LT} [ 1 + H(x)/Q^2 ]
\end{equation}
where $F_2^{LT}$ obeys the NLO QCD evolution equations and where $H(x)$
was parameterized as a fourth degree polynomial in $x$ (5 free
parameters). 

The normalizations of the ZEUS, H1 and NMC data were kept fixed 
to unity whereas those of the remaining data sets were allowed to float
within the quoted normalization errors (7 parameters). 

The uncertainties in the parton densities, structure functions and
related cross sections were estimated taking into account all
correlations. Included in the error calculation are the experimental
statistical errors, 57 independent sources of systematic uncertainty
(propagated using the technique described in~\cite{ref:pascaud}) and the
errors on the input parameters of the fit. 

\section{RESULTS}

The fit yielded a good description of the data with a $\chi^2$ of 1540 for
1578 data points and 28 free parameters. The quality of the fit is
illustrated in \Ab{fig:f2pvsq}
where we show the fixed target $F_2^p$
data for $x > 0.1$. 
\begin{figure}[htb]
\epsfig{file=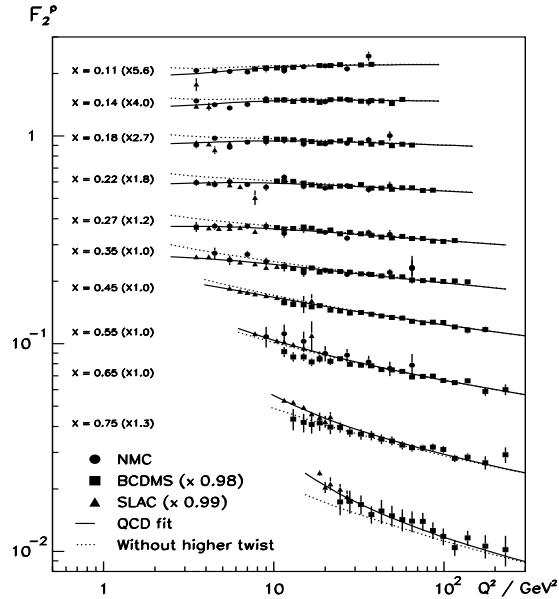,width=\linewidth,height=8cm,%
        bbllx=24,bblly=30,bburx=430,bbury=540}
\caption{
\footnotesize{
Fixed target data on $F_2^p$ versus $Q^2$ for $x > 0.1$. The full (dotted) curves
correspond to the QCD prediction with (without) higher twist contributions.
}
}
\label{fig:f2pvsq}
\end{figure}
The higher twist coefficient $H(x)$ in \Gl{eq:htwist} is found to be
negative for $x < 0.5$ and becomes large and positive at high $x$. Our
result on the higher twist contribution is very close to that obtained
by MRST~\cite{ref:mrsht}.

Nuclear effects in neutrino scattering were investigated by calculating
the ratios $xF_3^{\nu Fe}/xF_3^{\nu N}$ and $F_2^{\nu Fe}/F_2^{\nu N}$
of the CCFR data to the QCD predictions of scattering on a free
nucleon. Both these ratios do not significantly depend on $Q^2$ in
accordance with measurements of nuclear effects in muon
scattering~\cite{ref:nmcnuc}. The ratios, averaged over $Q^2$,
are plotted in \Ab{fig:nuceff} 
\begin{figure}[htb]
\begin{center}
\epsfig{file=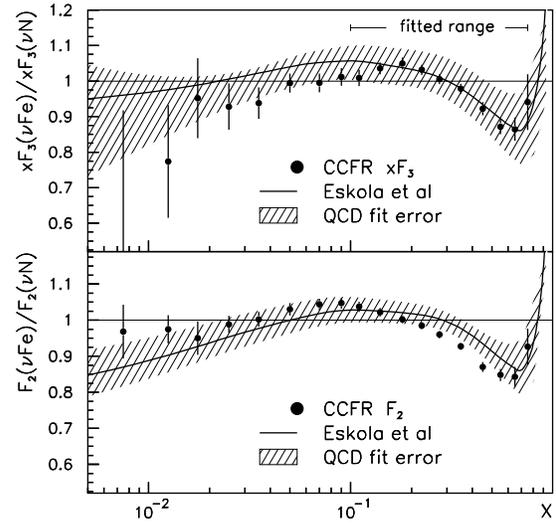,width=\linewidth,%
        bbllx=0,bblly=25,bburx=430,bbury=450}
\end{center}
\caption{
\footnotesize{
Nuclear effects in $\nu$-Fe scattering estimated from a comparison of
the CCFR data and the QCD fit as explained in the text.  The curves
correspond to the parameterizations from~\cite{ref:eskola} (used
to correct $xF_3^{\nu Fe}$ for $x > 0.1$). The errors from the QCD fit
are drawn as bands around the curves.
}
}
\label{fig:nuceff}
\end{figure}
and show the typical $x$ dependence of nuclear effects, including the
rise at large $x$ due to Fermi motion.  However, within the present
accuracy it is not possible to establish whether these nuclear effects
are significantly different from those measured in charged lepton
scattering (full curves in \Ab{fig:nuceff}). 

In \Ab{fig:allpdf} 
\begin{figure}[htb]
\begin{center}
\epsfig{file=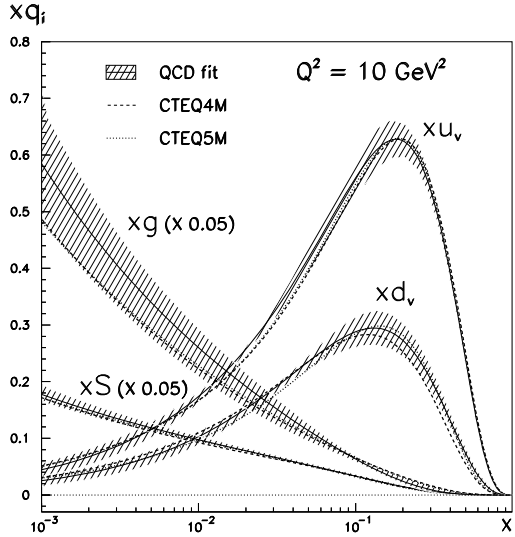,width=0.92\linewidth,%
        bbllx=20,bblly=30,bburx=520,bbury=550}
\end{center}
\caption{
\footnotesize{
The parton densities from this analysis compared to the results
from CTEQ4 and CTEQ5. The hatched bands indicate the error on
the QCD fit.
}
}
\label{fig:allpdf}
\end{figure}
are shown the parton densities obtained from the QCD fit (full curves).
We have verified that the strange quark density is compatible with
the measurements from CCFR~\cite{ref:ccfrs} and also that the QCD
prediction of $F_2^c$ agrees well with the ZEUS data~\cite{ref:zeusfc}
above $Q^2 = 10$~\gev.  This supports the assumption made in this
analysis that, at least for $x > 10^{-3}$, quark mass effects do not
spoil the QCD extrapolations to large $Q^2$. 

Also shown in \Ab{fig:allpdf} are the parton densities from CTEQ4 and
CTEQ5~\cite{ref:cteq5}. There is, within errors, good agreement between
these results and the QCD fit, although both the present analysis and
CTEQ5 yield a slightly harder $xd_v$ density than CTEQ4. Bodek and
Yang~\cite{ref:bodek} also obtain a harder $xd$ density by modifying
CTEQ4 (where the ratio $d/u \rightarrow 0$ as $x \rightarrow 1$) such
that
\begin{equation} \label{eq:bodek}
d/u \rightarrow d^{\prime}/u = d/u + Bx(1+x). 
\end{equation}
They find $B = 0.10 \pm 0.01$ which implies that $d/u \rightarrow 0.2$
as $x \rightarrow 1$. The $d/u$ ratio is shown in \Ab{fig:dou}.
\begin{figure}[htb]
\begin{center}
\epsfig{file=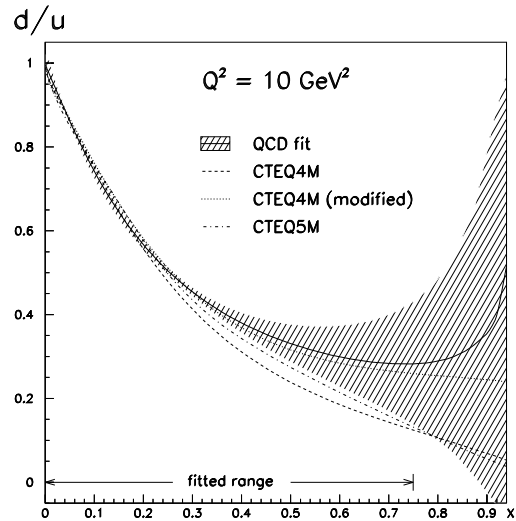,width=0.92\linewidth,%
        bbllx=20,bblly=30,bburx=530,bbury=550}
\end{center}
\caption{
\footnotesize{
The ratio $(d+\bar{d})/(u+\bar{u})$ from the QCD fit compared
to results from CTEQ4, CTEQ5 and
CTEQ4, modified according to \Gl{eq:bodek} with $B = 0.1$. 
The hatched band shows the error on the QCD fit.
}
}
\label{fig:dou}
\end{figure}
It is seen that the result from the QCD fit with $B = 0$ (full curve) is
for $x < 0.75$ close to the modified CTEQ4 distribution with $B = 0.1$
(dotted curve); if we leave $B$ a free parameter in the fit, we obtain
$B = -0.02 \pm 0.01$ (statistical error), close to zero. In any case,
the large error band clearly indicates that the exact behavior of $d/u$
at large $x$ is not well constrained. It might go to zero (CTEQ), to a
constant (Bodek and Yang) or may even diverge (this analysis) as $x
\rightarrow 1$. 

The CC $e^+p \rightarrow \bar{\nu}X$ cross section is predominantly
sensitive to 
the $d$ quark density 
\begin{equation}
{d^2 \si}/{dx dQ^2} \propto (1-y)^2(xd+xs)+x\bar{u}+x\bar{c}
\end{equation}
where $y = Q^2/(xs)$ with $s$ the $ep$ center of mass energy
($\sim$10$^5$~\gev\ at HERA). In \Ab{fig:ratios} 
\begin{figure}[htb]
\epsfig{file=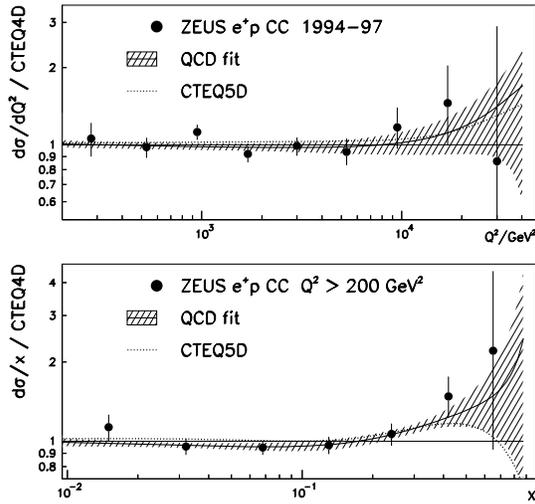,width=\linewidth,%
        bbllx=0,bblly=0,bburx=540,bbury=500}
\caption{
\footnotesize{
The CC $e^+p$ cross sections $d\si/dQ^2$ (top) and $d\si/dx$ (bottom)
measured by ZEUS, normalized to the QCD predictions based on CTEQ4.
The full (dotted) curves are the predictions from the QCD fit (CTEQ5).
The hatched bands show the error on the QCD fit prediction.
}
}
\label{fig:ratios}
\end{figure}
we show the ZEUS measurements~\cite{ref:ccxsec} of the CC $e^+p$ single
differential cross sections $d\si/dQ^2$ and $d\si/dx$ for $Q^2 >
200$~\gev, normalized to the NLO predictions calculated with the CTEQ4
parton distribution set. The full (dotted) curves correspond to the
predictions from the QCD fit (CTEQ5) which both achieve a better
description of the data at large $Q^2$ and $x$ due to an improved
determination of the $d$ quark density. SM predictions of the $e^+p$ NC
cross sections, calculated with the parton densities obtained in this
analysis, also agree very well with the recent ZEUS
data~\cite{ref:ncxsec} (not shown). 

We conclude that, within the present experimental accuracy, no
significant deviations can be observed between the data and the Standard
Model predictions.

\end{document}